\documentclass[aps,prl,twocolumn,superscriptaddress,amsfont,graphicx,nofootinbib,preprintnumbers]{revtex4}%
\usepackage{color,graphicx,epsfig}
\usepackage{ifpdf}
\usepackage{amsmath}
\usepackage{bm}
\usepackage{color}
\usepackage[english]{babel}
\usepackage{graphicx}%
\usepackage{amsfonts}%
\usepackage{amssymb}
\usepackage{braket}
\usepackage{hyperref}
\usepackage{enumerate}

\definecolor{nicered}{rgb}{0.7,0.1,0.1}
\definecolor{nicegreen}{rgb}{0.1,0.5,0.1}
\hypersetup{colorlinks,citecolor= nicegreen,linkcolor= nicered}

\newcommand{\ep}{\epsilon}

\newcommand{\be}{\begin{equation}}
\newcommand{\ee}{\end{equation}}
\newcommand{\bea}{\begin{eqnarray}}
\newcommand{\eea}{\end{eqnarray}}

\definecolor{Red}{rgb}{1.,0.,0.}

\def\dslash#1{\slash \!\!\!\! #1}

\def\OMIT#1{}

\begin{document}

\def\JHU{Department of Physics and Astronomy, Johns Hopkins University, Baltimore, USA}
\def\KIT{Institute for Theoretical Particle Physics, KIT, Karlsruhe, Germany}
\def\CERN{CERN Theory Division, CH-1211, Geneva 23, Switzerland}
\def\Argonne{High Energy Physics Division, Argonne National Laboratory, Argonne, IL 60439, USA}
\def\Northwestern{Department of Physics \& Astronomy, Northwestern University, Evanston, IL 60208, USA}

\preprint{CERN-PH-TH-2015-056}
\preprint{TTP15-017}

\title{Higgs boson production in association with a jet at next-to-next-to-leading order}

\author{Radja Boughezal}
\email[Electronic address:]{rboughezal@anl.gov}
\affiliation{\Argonne}

\author{Fabrizio Caola}            
\email[Electronic address:]{fabrizio.caola@cern.ch}
\affiliation{\CERN}

\author{Kirill Melnikov}            
\email[Electronic address:]{kirill.melnikov@kit.edu}
\affiliation{\KIT}

\author{Frank Petriello}     
\email[Electronic address:]{f-petriello@northwestern.edu}
\affiliation{\Argonne}
\affiliation{\Northwestern}

\author{Markus Schulze}            
\email[Electronic address:]{markus.schulze@cern.ch}
\affiliation{\CERN}

\begin{abstract}

We present precise predictions for Higgs boson production in association with a jet. Our calculation is accurate to next-to-next-to-leading order (NNLO) QCD in 
the Higgs Effective Field Theory and constitutes 
the  first complete NNLO computation for Higgs production with a final-state jet in hadronic collisions. 
We include all relevant phenomenological channels and present fully-differential results as well as total cross sections for the LHC. 
Our NNLO predictions reduce the unphysical scale dependence by more than a factor of two and enhance the total rate by about twenty percent compared to NLO QCD predictions. 
Our results demonstrate for the first time satisfactory convergence of the perturbative series.

\end{abstract}

\maketitle

Further  exploration of the Higgs boson discovered by the ATLAS and CMS collaborations~\cite{:2012gk,:2012gu} will be a primary goal 
of the continued experimental program of the LHC.  
In the well-measured decay modes, $h\to \gamma\gamma$, $WW$ and $ZZ$,  the determined couplings agree with 
the Standard Model (SM) values to  $20-30$ percent~\cite{atlascoup,cmscoup}.  This agreement will be further probed during the upcoming LHC run.  
The predictions of the SM are expected to be tested to the five percent  level in several production and decay modes~\cite{Dawson:2013bba}, providing an exciting opportunity
to discriminate between different mechanisms of electroweak symmetry breaking.
In addition, new properties of the Higgs boson will be accessed through the 
measurement of its kinematic  distributions.  These measurements will test whether the tensor structures of the Higgs couplings are correctly 
predicted by the SM, whether additional operators mediate Higgs production and decay, and whether there are new  
particles that modify the loop-induced $ggH$ and $\gamma \gamma H$ couplings.  

Such studies~\cite{atlasdiff}
are currently limited by the available statistics.  However, this situation will change during Run II of the LHC, and   
eventually the limiting factor in the search for deviations in Higgs boson properties from predictions will be our understanding of SM theory.  
This is apparent from a recent coupling combination performed by ATLAS~\cite{atlaserrors}.  The uncertainty on the theoretical predictions dominates 
the systematic error budget in all of the di-boson decay modes.  One component of this error is  the overall signal normalization, 
for which a precise calculation of inclusive Higgs production in the gluon-fusion channel is needed.  
The completion of the next-to-next-to-next-to-leading order (N$^3$LO) calculation was recently announced~\cite{nnnlo}.  
The other major component of the theoretical error is the need for improved predictions of the differential spectra that enter every analysis.  
In some final states this need is obvious; for example, in the di-leptonic decay of the $WW$ channel a mass peak cannot be reconstructed, and 
accurate calculations of both signal and background distributions are needed in order to devise appropriate experimental search strategies.  
Even in modes where a resonance peak can be reconstructed, such as the $\gamma\gamma$ channel, the Higgs candidate events are categorized according 
to their  transverse momentum ($p_\perp$) in order to improve the signal significance.  
Such a division relies upon accurate and precise theoretical predictions for the Higgs $p_\perp$ and rapidity distributions,
that are used to reweight the parton-shower Monte Carlo simulations employed  by the experimental collaborations. 

In this Letter we take a major step toward improving SM predictions for several kinematic distributions 
employed in the analysis of Higgs boson properties, 
by providing a next-to-next-to-leading order (NNLO) 
calculation of Higgs boson 
production in association with a jet. Compared to previous computations~\cite{Boughezal:2013uia,Chen:2014gva}, we include all relevant channels and color structures.
  The phenomenological impact of this result spans all Higgs 
search channels.  In the $WW$ final state it refines the division of the signal prediction into exclusive zero-jet, 
one-jet and inclusive two-jet bins, and it can be used to improve the resummation 
of the jet-veto logarithms that accompany this division~\cite{jvlogs}.  
For all final states our calculation  can be used to more accurately re-weight the Higgs $p_\perp$ distribution obtained from Monte Carlo.  
Finally,  it will allow for the comparison of the measured differential distributions from LHC Run II with the most precise SM theory to more incisively probe the mechanism of electroweak symmetry breaking.  

Our calculation also represents a technical milestone in the application of perturbative QCD to the modelling of hadronic collisions.
The past few years have seen a renaissance in the development of subtraction techniques designed to turn our knowledge of the infrared structure 
of QCD at NNLO into actual phenomenological predictions for hadron-collider observables~\cite{nnlosub,Boughezal:2013uia,Chen:2014gva}.  Our result demonstrates the power of 
these newly-developed methods in assisting the continued exploration of Nature at the LHC.

Our Letter  is organized as follows.  We first review the theoretical framework that we use to obtain the results reported in this paper.  
Since this has 
been described in detail in our previous work on Higgs plus jet production in pure gluodynamics~\cite{Boughezal:2013uia}, we only sketch here the 
salient features of the calculation.  We then present the numerical results of the computation including NNLO results for cross sections 
of Higgs plus jet production at various collider energies and for various values of the transverse momentum cut on the jet. We also discuss the NNLO QCD 
corrections to the transverse  momentum distribution of the Higgs boson.  Finally, we present our conclusions. 

\begin{figure}[t]
 \centering
 \includegraphics[width=1\columnwidth]{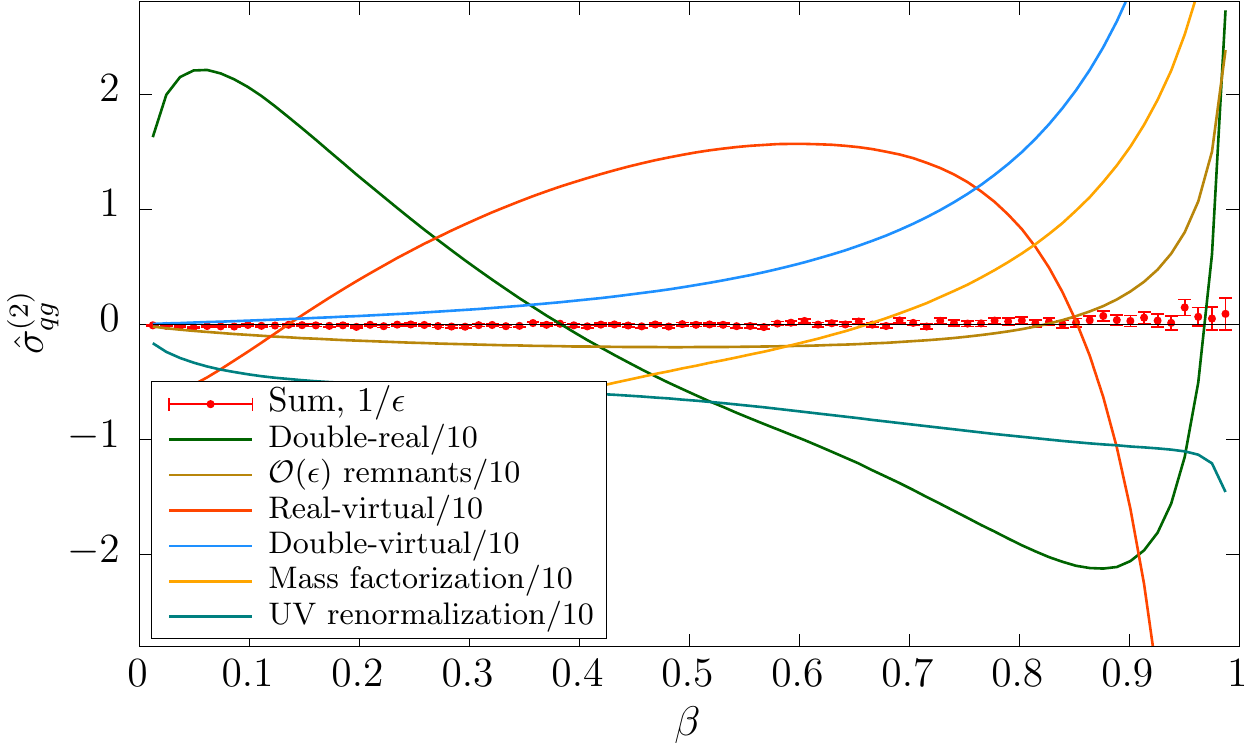}
 \caption{Cancellation of $1/\ep$ poles in the $qg$ channel. Note that individual contributions have been rescaled by a factor of $0.1$, while the sum of them
   is not rescaled.}
 \label{fig1}
\end{figure}

 
We begin by reviewing the details of the computation. Our calculation  is based on the effective theory obtained by integrating out the 
top quark.  For values of the Higgs $p_\perp$ below 150 GeV, this approximation is known to work to 3\% or better at NLO~\cite{Harlander:2012hf,Dawson:2014ora}.
Since the Higgs boson receives its transverse momentum by recoiling against jets, we expect that a similar accuracy of the 
large-$m_t$ approximation can be expected for observables where jet transverse momenta do not exceed ${\cal O}(150)~{\rm GeV}$ as well.

The effective Lagrangian is given by 
\be
{\cal L} = -\frac{1}{4} G_{\mu \nu}^{(a)} G^{(a),\mu \nu} +\sum_i \bar{q}_i {\mathrm i} \dslash{D} q_i 
- C_1 \frac{H}{v} G_{\mu \nu}^{(a)} G^{(a),\mu \nu},
\label{eq:lag1}
\ee
where $G_{\mu \nu}^{(a)}$ is the gluon field-strength tensor,  $H$ is the Higgs boson field and  $q_i$ denotes 
the light quark field of flavor $i$.  The flavor index runs over the values $i=u,d,s,c,b$, which are all taken to be massless.  
The covariant derivative $\dslash{D}$ contains the quark-gluon coupling. The Higgs vacuum expectation value is denoted by   
$v$, and $C_1$ is the Wilson coefficient obtained by integrating out the top quark.  The calculation presented here 
requires $C_1$ through ${\cal O}(\alpha_s^3)$, which can be obtained from Ref.~\cite{Chetyrkin:1997un}.  
Both the Wilson coefficient and the strong coupling constant require ultraviolet renormalization; the corresponding renormalization 
constants can be found e.g. in Ref.~\cite{Anastasiou:2002yz}.   
 
Partonic cross sections computed according to the above prescription are still not finite physical quantities.  Two remaining issues must be addressed.  First, 
contributions of final states with different number of  partons must be combined in  an appropriate way to produce  infrared-safe observables.  This requires a
definition of final states with jets.  We use the anti-$k_T$ jet algorithm~\cite{Cacciari:2008gp} to combine partons into jets. 
Second, initial-state collinear singularities must be absorbed into the parton distribution 
functions (PDFs) by means of standard $\overline {\rm MS}$ PDF renormalization. A detailed 
discussion of this procedure can be found in Ref.~\cite{Buehler:2013fha}.

\begin{figure}[t]
 \centering
 \includegraphics[width=1\columnwidth]{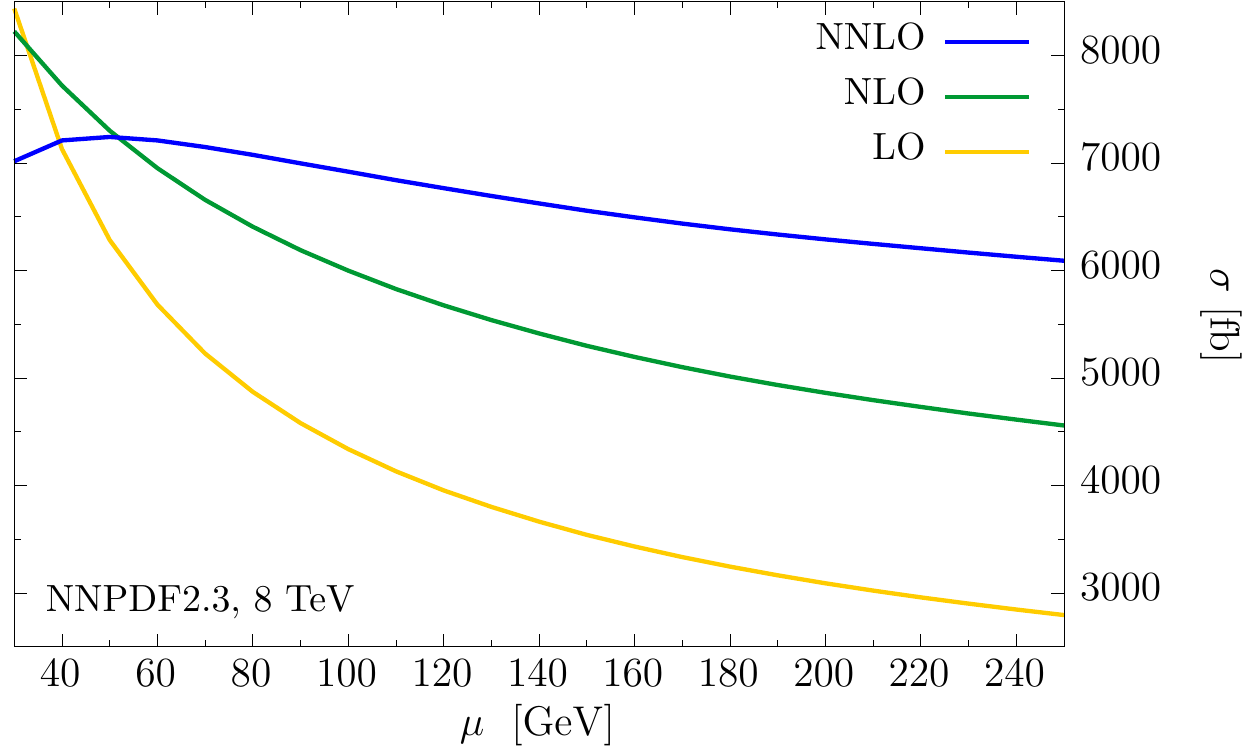}
 \caption{ Dependence of the total LO, LO and NNLO cross-sections on the unphysical scale $\mu$. See text for details.}
 \label{fig2}
\end{figure}

The finite cross sections for each of the partonic channels $ij$ obtained in this way have an 
expansion in the $\overline {\rm MS}$ 
strong coupling constant $\alpha_s \equiv \alpha_s(\mu)$, defined in a theory with five active flavors, 
\be
\sigma_{ij} = \sigma^{(0)}_{ij} + \frac{\alpha_s}{2\pi} \sigma^{(1)}_{ij} + \left(\frac{\alpha_s}{2\pi}\right)^2 \sigma^{(2)}_{ij} + {\cal O}(\alpha_s^6).
\ee
Here, the omitted terms indicated by ${\cal O}(\alpha_s^6)$ include the $\alpha_s^3$ factor that is contained in the leading order cross section 
$\sigma^{(0)}_{ij}$.   Our 
computation will include the $gg$ and $q g$ partonic cross sections at NNLO, $\sigma^{(2)}_{gg}$ and $\sigma^{(2)}_{qg}$, where  
$q$ denotes any light quark or anti-quark.  At NLO, it can be checked using MCFM~\cite{Campbell:2010ff} that these  channels contribute 
over 99\% of the cross section for typical jet transverse momentum cuts, $p_\perp \sim 30$ GeV.  We therefore include the partonic  channels 
with two quarks or anti-quarks in the initial state only through NLO.

In addition to the ultraviolet and collinear renormalizations described above, we need the following 
ingredients to determine  $\sigma^{(2)}_{gg}$ and $\sigma^{(2)}_{qg}$: the two-loop virtual corrections to the partonic channels $gg \to Hg$ and $qg \to Hq$; 
the one-loop virtual corrections to $gg \to Hgg$, $gg \to Hq\bar{q}$ and $qg \to Hqg$; the double real emission processes 
$gg \to Hggg$, $gg \to Hgq\bar{q}$, $qg \to H qgg$ and $qg \to HqQ\bar{Q}$, where the $Q\bar{Q}$  pair in the 
last process can be of any flavor.  The helicity amplitudes for 
all of these processes are available in the literature. The 
two-loop amplitudes were computed  in  Ref.~\cite{Gehrmann:2011aa}. 
The one-loop corrections to the four-parton processes are known~\cite{gghgg_1loop} and available as a Fortran code in the MCFM program~\cite{Campbell:2010ff}. 
For five-parton tree-level amplitudes, we use compact results obtained using BCFW recursions~\cite{simon_tree}.

The difficulty 
in completing the NNLO calculation becomes apparent when one attempts to combine these contributions and cancel the infrared divergences that appear separately 
in each component.  The problem is that final states with different multiplicities live  in different phase-spaces; this feature makes it impossible 
to combine them directly. The issue becomes obvious if one looks at how $1/\ep$ singularities appear in different contributions. 
Indeed,  the $1/\ep$ poles coming from loop amplitudes are explicit ones, but those coming from the real-emission corrections only appear upon 
integration over the unresolved region of phase space.  However, since we want to keep the calculation fully differential, 
we want to avoid integrating over the phase-space for higher-multiplicity processes.  

To reconcile these two requirements, which at first sight appear to be mutually 
exclusive, we use the sector-improved residue subtraction approach~\cite{Czakon:2010td,Czakon:2011ve,Boughezal:2011jf,Czakon:2014oma}.  This is an outgrowth 
of the sector-decomposition method~\cite{Binoth:2003ak,Anastasiou:2003gr,Binoth:2004jv} used to compute the differential cross sections for Higgs boson and electroweak gauge bosons through NNLO.  Sector decomposition uses the observation that the relevant singularities can be isolated 
using appropriate parameterizations of phase space and expansions in plus distributions.  Sector-improved residue subtraction combines this with the idea that a pre-partitioning of the final-state phase space similar to the FKS subtraction used at NLO~\cite{Frixione:1995ms} 
allows to  extend this technique 
to $2 \to 2$ and more complicated scattering processes.  A detailed discussion of the 
phase-space parameterizations needed to handle all the contributing partonic processes was given in Ref.~\cite{Boughezal:2013uia}, to which we refer the 
reader for more details.  Note, however, that Ref.~\cite{Boughezal:2013uia} dealt with $gg \to H+ng$ partonic processes for which  both the phase-space and 
the matrix elements are highly symmetric. For the quark-gluon channel this symmetry is lost and one has to consider a larger number of ``sectors'' compared 
to the case of pure gluodynamics. 

Before discussing numerical results, we would like  to point out  two things in connection with the application of 
sector-improved residue subtraction  method. 
First, we note that upon applying this method, one automatically generates subtraction terms that allow  extraction 
of $1/\ep$ singularities and, at the same time,  make integration of the finite remainders possible. 
The key point is that these subtraction terms are obtained from universal soft and collinear limits of scattering amplitudes that 
were computed long ago in Refs.~\cite{lcat1,lb, lmc3, lcat2, lzb1, lcat3, lcat4,Kosower:1999rx}.  The universality of these subtraction 
terms makes the method of improved-sector decomposition attractive and, in principle, applicable to processes of arbitrarily high multiplicity. 
Second, when sector-improved residue subtraction is applied to a physical process, it leads to a Laurent expansion 
of the various contributions to  the cross section in the  dimensional regularization parameter $\epsilon$; the coefficients of this expansion are computed {\it numerically}. Since 
final physical cross sections are independent of the regularization parameter, the quality of $1/\epsilon^n$, $n=4,3,2,1$, cancellation is an important check 
of the correctness of the implementation of the method. To show the quality of the cancellation in our case, in Fig.~\ref{fig1} we present  various 
contributions to the $1/\epsilon$ pole of the partonic cross section, together with the residual non-cancellation, in the $qg$ channel.
We show these quantities as functions of the distance from the partonic threshold, defined  as $\beta=\sqrt{1-s_{th}/\hat s},~~~\sqrt{s_{th}}=\sqrt{m_H^2+p_{\perp,cut}^2} + p_{\perp,cut}$. 
We see that the cancellation is very good, at the level of one per mill 
or better.   Although in Fig.~\ref{fig1} we display the result for the total cross section, we have also checked that the cancellation holds at a similar 
level for kinematic distributions. 

In addition,  we have checked that our results for the Higgs plus 2-jet cross 
section at NLO agree with  MCFM~\cite{Campbell:2006xx}, 
for both the fiducial cross section and for several kinematic distributions.  We have two separate 
numerical implementations of the sector-improved residue subtraction method that demonstrate good agreement.  
Furthermore, an independent calculation was also performed using the jettiness-subtraction technique~\cite{FrankRadjaHpJ}, 
and good agreement for the fiducial cross sections was  found.

 
We now turn to the discussion of  numerical results.   We first compute the LO, NLO and NNLO cross 
sections for Higgs plus jet production $pp \to H+j$ 
at the $8$ TeV and $13$ TeV LHC. We use $m_H=125~{\rm GeV}$ and $m_t=172.5~{\rm GeV}$. To define the cross section, 
we use the anti-$k_\perp$ algorithm with $\Delta R = 0.5$ and 
a cut on the jet transverse momentum 
$p_\perp > 30~{\rm GeV}$.   We employ parton distribution functions and the strong coupling constant as provided by the NNPDF21LO~\cite{Ball:2011mu}, NNPDF23NLO and NNPDF23NNLO~\cite{Ball:2012cx} 
PDF sets to compute respectively LO, NLO and NNLO cross sections.  We set the renormalization and factorization scales to the mass 
of the Higgs boson and we estimate the uncertainty associated with higher orders in perturbation theory by changing the scale 
by a factor of two. For the $8$~TeV LHC, we find  $ \sigma_{pp \to H+j} =   3.9^{+1.7}_{-1.1}~{\rm pb},\; 5.6^{+1.3}_{-1.1}~{\rm pb},\;  6.7^{+0.5}_{-0.6}~{\rm pb}$ at leading, next-to-leading 
and next-to-next-to-leading order, respectively. Results for $\mu=m_H/2$ and $\mu=2 m_H$ are shown as super- and sub-scripts, respectively.
 For $\mu = m_H$, the NLO (NNLO)  cross section exceeds the leading order one by $44\% \; (72\%)$, indicating reasonable convergence of perturbative expansion. The convergence is better
for lower scales: for example, for $\mu = m_H/2$ the NLO (NNLO) cross section exceeds the leading order one by $23\% \; (29\%)$. As expected, the scale uncertainty is significantly
reduced at NNLO. This is also illustrated in Fig.~\ref{fig2}, where we plot the total cross section at LO, NLO and NNLO as a function of the unphysical scale $\mu$ over the range
$\mu \in [p_{\perp,cut}:2 m_H]$. We estimate the residual uncertainty due to PDF to be at the $\mathcal O(5\%)$ level. 
The situation is similar for the 13 TeV LHC. More precisely, we find $\sigma_{pp\to H+j}= 10.2^{+4.0}_{-2.6}~{\rm pb},\; 14.7^{+3.0}_{-2.5}~{\rm pb},\; 17.5^{+1.1}_{-1.4}~{\rm pb}$ at leading, next-to-leading
and next-to-next-to-leading order, corresponding to a NLO (NNLO) increase with respect to LO of $44\% \; (72\%)$ for $\mu=m_H$ and of $25\% \; (31\%)$ for $\mu=m_H/2$.

\begin{figure}[t]
 \centering
 \includegraphics[width=1\columnwidth]{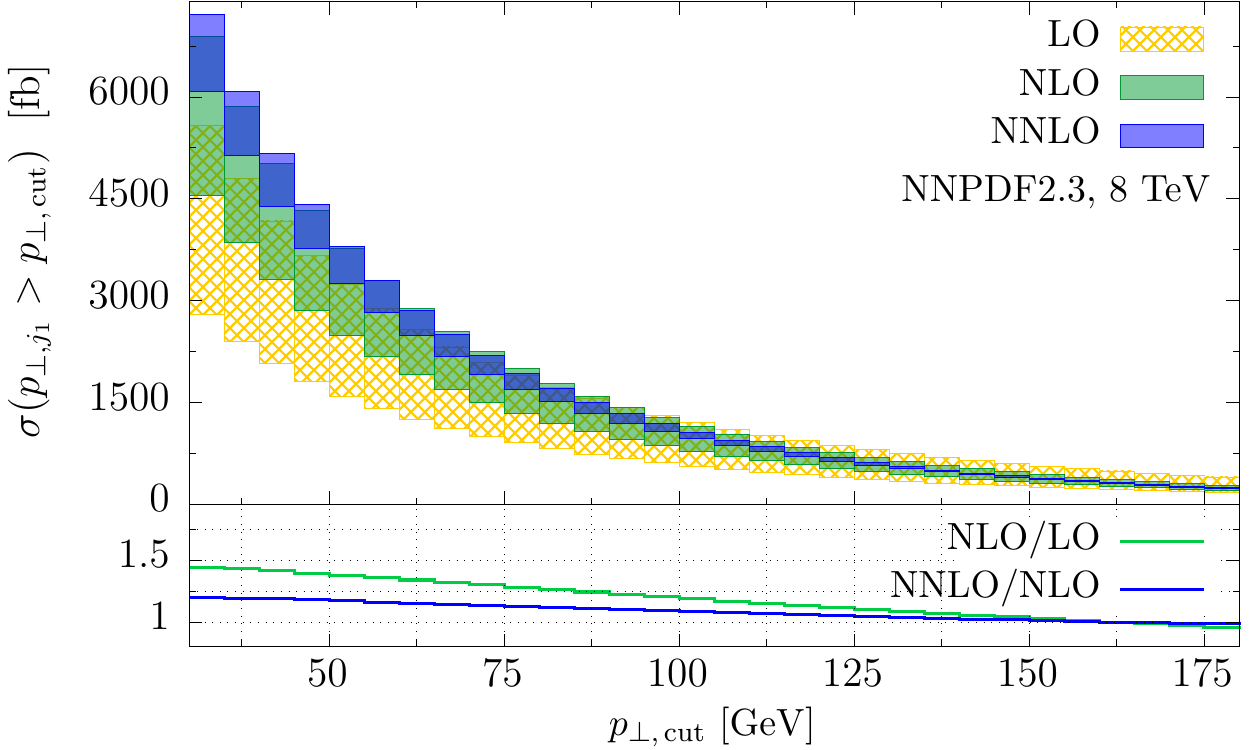}
 \caption{Higgs plus jet production cross-sections in dependence of the cut on the jet transverse momentum. The minimal cut we consider is
$p_\perp > 30~{\rm GeV}$. See text for details.}
 \label{fig3}
\end{figure}

It is interesting   to understand to what extent perturbative  QCD corrections depend on the 
kinematics of the process and/or on the details of the jet algorithm.
One way to study this is to explore how the 
NNLO QCD corrections change as the lower cut on the jet transverse momentum is varied. 
We show corresponding results for the 8 TeV LHC in Fig.~\ref{fig3} where the cumulative distribution for  $\sigma(H+j,p_{\perp,j} \ge p_{\perp,{\rm cut}})$ is displayed. 
The inset in Fig.~\ref{fig3} shows  ratios of  NNLO(NLO) to NLO(LO) $H+j$  cross-sections, respectively,  
computed  for $\mu_F = \mu_R = m_H$ as a function of the jet $p_\perp$-cut.
It follows from Fig.~\ref{fig3}  that QCD radiative corrections depend on the kinematics. Indeed, 
the NNLO to NLO cross-sections ratio changes from $1.25$ at $p_\perp = 30~{\rm GeV}$ to $\sim 1$ at $p_\perp \sim 150~{\rm GeV}$. 

\begin{figure}[t]
 \centering
 \includegraphics[width=1\columnwidth]{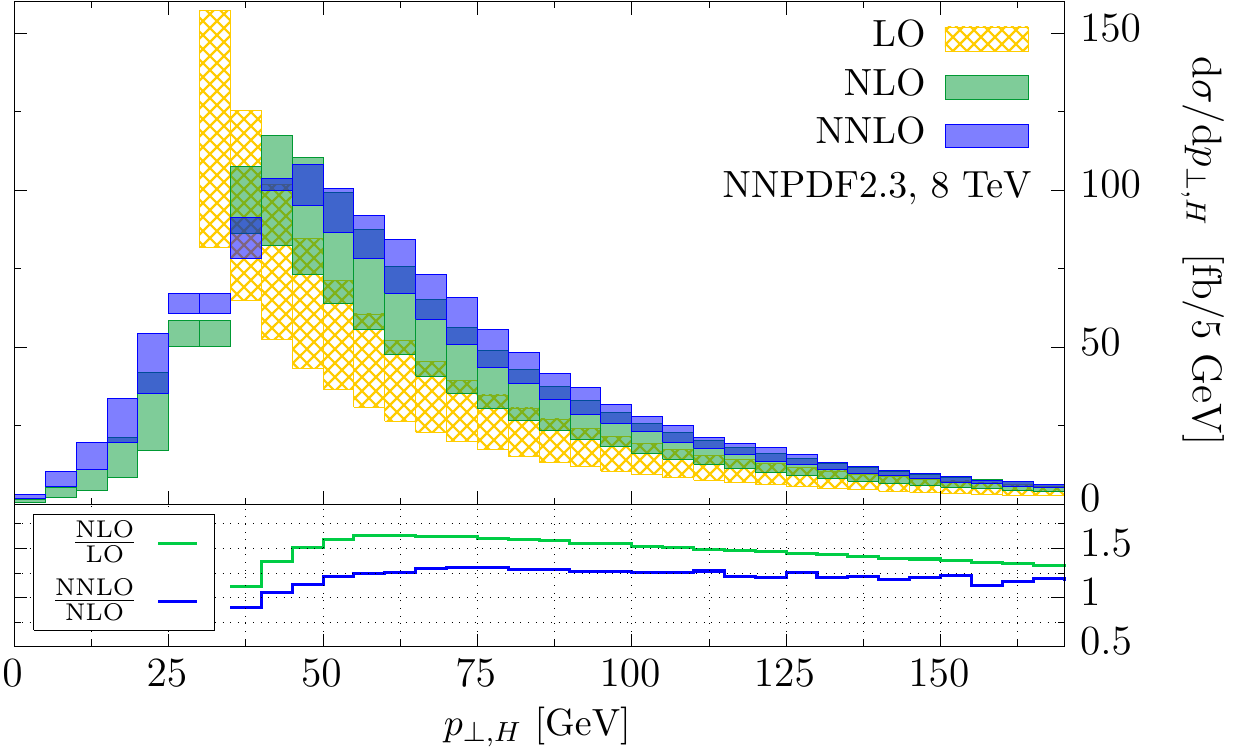}
 \caption{ Higgs boson transverse momentum distribution in $pp \to H+j$ at $8$ TeV LHC. The jet is defined
with the anti-$k_\perp$ algorithm with $\Delta R = 0.5$ and the cut on the jet transverse momentum of $30~{\rm GeV}$.
Further details are explained in the  text.}
 \label{fig4}
\end{figure}

In Fig.~\ref{fig4} we show the Higgs boson transverse momentum distribution in the reaction $pp \to H+j$, 
for three consecutive orders of perturbation theory.  We require that there 
is a jet in the final state with a transverse momentum higher than $p_{\perp,j} > 30~{\rm GeV}$.  
Note that the two bins closest to the boundary $p_{\perp,H}=30$ GeV have been combined to avoid the well-known Sudakov-shoulder effect~\cite{Catani:1997xc}.
Away from that region, the NNLO QCD radiative corrections increase the NLO cross-section by about $20\%$, 
slowly decreasing as $p_{\perp,H}$ increases.

In conclusion, we have presented a calculation of the NNLO QCD corrections to the production of the Higgs boson in association with a jet at 
the LHC. This is the first complete computation of NNLO QCD corrections to a Higgs production process with a jet in the final state. 
It shows that techniques for performing NNLO QCD  computations,  that were in the development phase for several years, can indeed be used to provide 
precise predictions for complex process at hadron colliders.   The total cross section for $H+$jet production receives moderate NNLO QCD 
corrections. For jets defined with the anti-$k_\perp$ algorithm with $p_{\perp, j} > 30~{\rm GeV}$, we find NNLO QCD corrections 
 of the order of $20\%$ for $\mu = m_H$.  These moderate corrections are the result of the smaller corrections for the $qg$ channel w.r.t the $gg$ one,
and a suppression of the $gg$ channel due to $q\bar q$ final states not considered in previous analyses~\cite{Boughezal:2013uia,Chen:2014gva}.
Beyond the total cross section, our computation will have important implications for many processes  that are used 
to study properties of the Higgs boson, including $W^+W^-$ and $\gamma \gamma$ final states, primarily through improved modelling of the Higgs 
transverse momentum and rapidity distributions.  In particular, since the complete N$^3$LO computation of the Higgs boson production cross section is available, 
a consistent computation of the $H+0$~jets, $H+1$~jet, $H+2$~jet and $H+3$~jet exclusive processes becomes  possible for the first time.  
Furthermore, since the Higgs boson is a spin-zero particle, our computation can be easily extended to include Higgs boson decays, to enable 
theoretical predictions for fiducial cross sections and kinematic distributions for the particles that are observed in detectors.  
Once this is done, our calculation will 
provide a powerful tool that will help to understand detailed properties of the Higgs boson at the LHC. 

We thank T.~Becher, J.~Campbell, T.~Gehrmann and M.~Jaquier for helpful communications. 
We are grateful to S.~Badger for making his results for tree-level amplitudes available to us.
F.~C. would like to thank the Institute for Theoretical Particle Physics of KIT and the Physics and Astronomy Department of Northwestern University for hospitality at various stages of this project.
R.~B. is supported by the DOE under the contract DE-AC02-06CH11357.  F.~P. is supported by the DOE grants DE-FG02-91ER40684 and DE-AC02-06CH11357.  
This research used resources of the National Energy Research Scientific Computing Center, a DOE Office of Science User Facility supported by the Office of Science 
of the U.S. Department of Energy under Contract No. DE-AC02-05CH11231.

\end{document}